\documentclass{svproc}
\usepackage{url}
\usepackage{graphicx}

\usepackage{hyperref}

\begin{document}
\mainmatter              
\title{Gamma Strength Functions and the Brink-Axel Hypothesis}
\titlerunning{Gamma Strength Functions and the Brink-Axel Hypothesis}  
%
\author{Peter von Neumann-Cosel}
\authorrunning{Peter von Neumann-Cosel} 
\institute{Institut f\"ur Kernphysik, Technische Universit\"at Darmstadt, 64289 Darmstadt, Germany\\
\email{vnc@ikp.tu-darmstadt.de}}

\maketitle              
\begin{abstract}

Experimental tests of the Brink-Axel hypothesis relating gamma strength functions (GSF) deduced from absorption and emission experiments are discussed.
High-resolution inelastic proton scattering at energies of a few hundred MeV  and at very forwrd angles including $0^\circ$ presents a new approach to test the validity of the BA hypothesis in the energy region of the pygmy dipole resonance. 
Such data not only provide the GSF but also the level density (LD) and thus permit an independent test of their model-dependent decomposition in the Oslo method. 
\end{abstract}

\section{Gamma Strength Function}
\begin{figure}[tbh]
\begin{center}
\includegraphics[width=7cm]{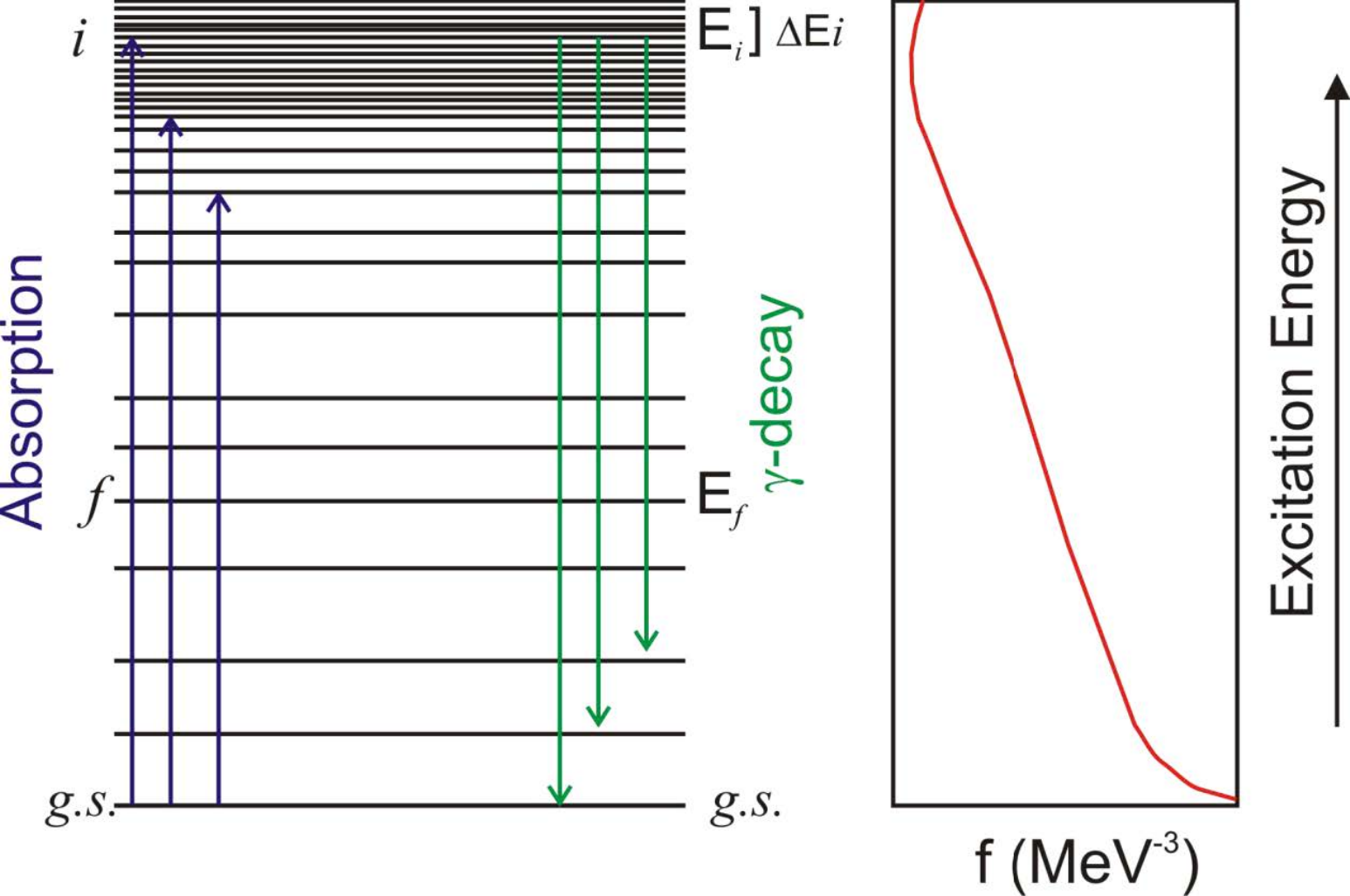}
\caption{\label{fig1}
Relation between $\gamma$ decay and absoprtion (l.h.s.) and expected energy dependence of the GSF (r.h.s.). }
\end{center}
\end{figure}

The GSF describes the average $\gamma$ decay behavior of a nucleus.
It depends on the level densities at the initial and final energies.
In general all multipoles allowed for electromagnetic processes contribute but in practise E1 dominates. 
Thus, the isovector giant dipole resonance (IVGDR) dominates the GSF at higher excitation energies as indicated on the r.h.s.\ of Fig.~\ref{fig1}. 
At lower energies M1 contributes to the total GSF  (although a few \% only under most conditions).

As indicated in the scheme of decay and absorption in Fig.~\ref{fig1}, for the special case of $\gamma$ decay to the g.s.\ the GSF can be related to the photoabsorption cross section by the principle of detailed balance
\begin{equation}
	\label{eq1}
	f^{E1}(E_\gamma, J) = \frac{2J_0+1}{2J+1}\frac{1}{(\pi\hbar c)^2E_\gamma^3}
	\left\langle\sigma_{abs}\right\rangle
\end{equation}
where $J, J_0$ are the spins of excited and ground state, respectively, and for simplicity the relation is written for the $E1$ component only.
The brackets $\langle \rangle$ indicate averaging over an energy interval.

\section{Experimental Tests of the Brink-Axel Hypothesis}
Knowledge of the GSF is required for calculations of statistical nuclear reaction in astrophysics \cite{arn07}, reactor design \cite{cha11}, and waste transmutation \cite{sal11}.
Most applications imply an environment of finite temperature, notably in stellar scenarios \cite{wie12}, and thus reactions on excited states (e.g.\ in a (n,$\gamma$) reaction) become relevant.
Their contributions to the reaction rates are usually estimated applying the generalized Brink-Axel (BA) hypothesis \cite{bri55,axe62}, which states that the GSF is independent of the properties of the initial and final states (and thus should be the same in $\gamma$ emission and absorption experiments).
Although historically formulated for the IVGDR, where it seems to hold approximately for not too high temperatures \cite{bbb98}, this is nowadays a commonly used assumption to calculate the low-energy E1 and M1 strength functions. 
Recent theoretical studies \cite{joh15,hun17} put that into question demonstrating that the strength functions of collective modes built on excited states do show an energy dependence.
However, numerical results for E1 strength functions  showed an approximate constancy consistent with the BA hypothesis \cite{joh15}. 

The so-called Oslo method, where primary spectra of $\gamma$ decay following compound nuclear reactions are extracted, is a major source of data on the GSF below the particle thresholds.
Since the $\gamma$ transmission probability is proportional to the product of the GSF and the final-state LD, assumption of the generalized BA hypothesis is a prerequisite of the analysis \cite{sch00}.     
Recent Oslo-type experiments have indeed demonstrated independence of the GSF from excitation energies and spins of initial and final states in a given nucleus in accordance with the BA hypothesis \cite{gut16,lar17}.
However, there are a number of results which clearly indicate violations in the low-energy region when comparing $\gamma$ emission and absorption experiments.
For example, the GSF in heavy deformed nuclei at excitation energies of $2 -3$ MeV is dominated by the orbital M1 scissors mode \cite{boh84} and potentially large differences in B(M1) strengths are observed between $\gamma$ between upward \cite{hey10} and downward \cite{gut12,ang16} GSFs. 
Furthermore, at very low energies ($< 2$ MeV) an increase of GSFs is observed in Oslo-type experiments \cite{lar17,voi04}, which for even-even nuclei cannot have a counterpart in ground state absorption experiments on even-even nuclei because of the pairing gap.           

For the low-energy E1 strength in the region of the PDR, the validity of the BA hypothesis is far from clear when comparing results from the Oslo method with photoabsorption data.
Below particle thresholds most information on the GSF stems from nuclear resonance fluorescence (NRF) experiments, which suffers from the problem of unobserved braching ratios to excited states. 
These can be corrected in principle by Hauser-Feshbach calculations assuming statistical decay \cite{rus09}.
The resulting correction factors are sizable and show a strong dependence on the neutron threshold energy and the g.s.\ deformation.
On the other hand, there are clear indications of non-statistical decay behavior of the PDR from recent measurements \cite{rom15,loe16,isa19}.
Violation of the BA hypothesis was also claimed in a simultaneous study of the $(\gamma,\gamma^\prime)$ reaction and average ground state branching ratios \cite{ang12} in $^{142}$Nd (see, however, Ref.~\cite{ang15}).   
Clearly, information on the GSF in the PDR energy region from independent experiments is called for.

\section{GSF and LD from (p,p$^\prime$) scattering}

A new method for the measurement of complete E1 strength distributions in nuclei from about 5 to 25 MeV has been developed using relativistic Coulomb excitation in inelastic proton scattering at beam energies of a few hundred MeV and  scattering angles close to $0^\circ$ \cite{tam11,pol12,kru15,has15,bir17,mar17}.
The experiments also permit extraction of the M1 part of the GSF due to spinflip excitations \cite{bir16}, which energetically overlaps with the PDR strength.
Furthermore, when performed with good energy resolution, the level LD can be extracted independently of the GSF in the excitation region of the IVGDR from the giant resonance fine structure \cite{pol14}.
This allows an important test of the model-dependent decomposition of LD and GSF in the Oslo method \cite{sch00} . 

\begin{figure}[t]
\begin{center}
\includegraphics[width=11cm]{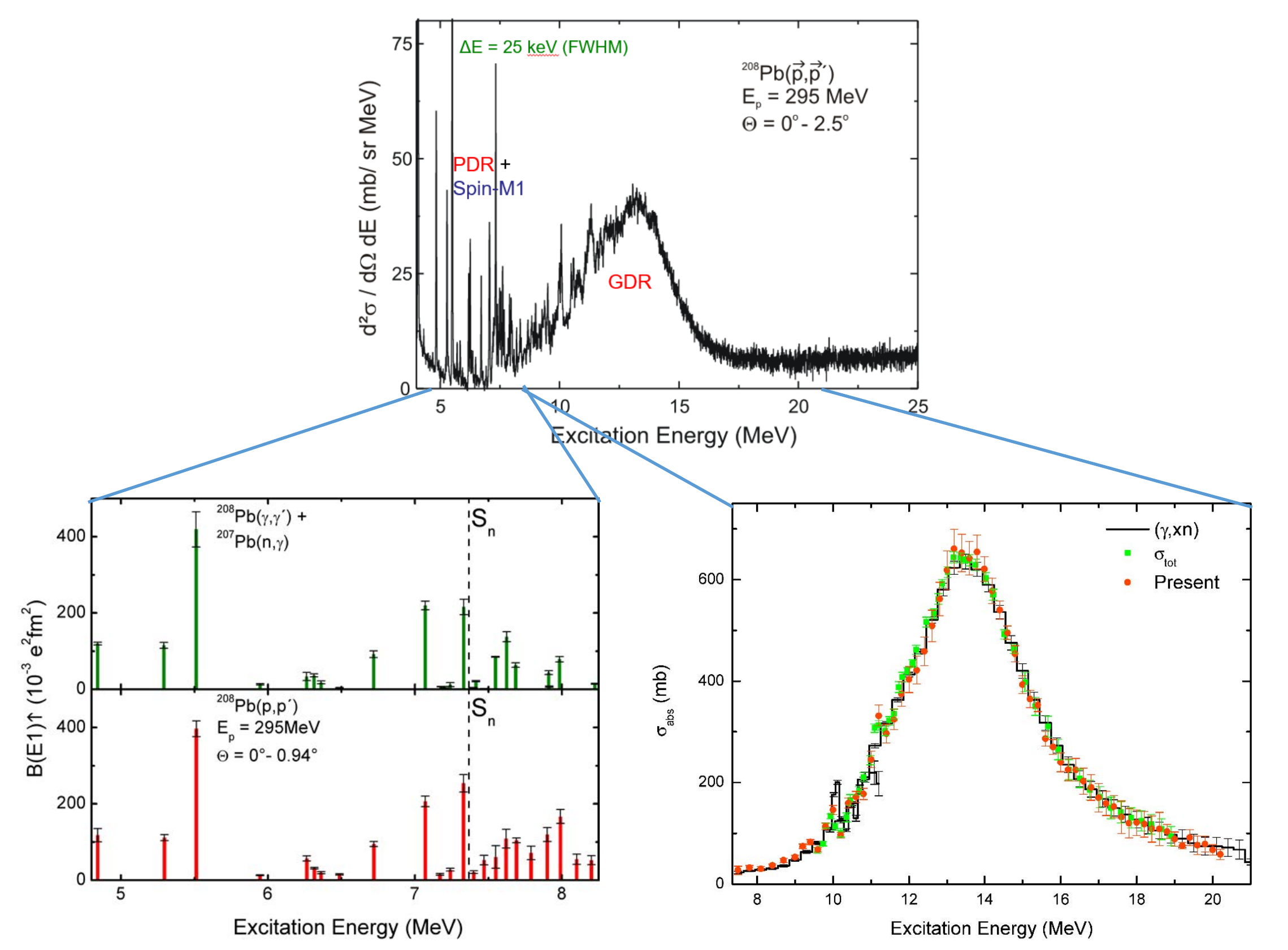}
\caption{\label{fig2}
Top: Experimental spectrum of the $^{208}$Pb(p,p$^\prime$) reaction at $E_0 = 295$ MeV and $\Theta_{\rm lab} = 0^\circ$.
Bottom: Comparison of the B(E1) strength distribution deduced from the (p,p$^\prime$) experiment below $S_{\rm n}$ and the photoabsorption cross section above $S_{\rm n}$ with results from other experiments.
See Refs.~\cite{tam11,pol12}.}
 \end{center}
\end{figure}

The case of $^{208}$Pb is used as an example to illustrate the methods \cite{tam11,pol12} and the comparison to the Oslo data \cite{bas16}.
Details on the experimental techniques can be found in Ref.~\cite{tam09}. 
The top part of Fig.~\ref{fig2} shows a spectrum of the  $^{208}$Pb(p,p$^\prime$) reaction in the excitation region $E_{\rm x} = 4 - 25$ MeV measured with the magnetic spectrometer placed at $0^\circ$.
One observes prominent transitions at low excitation energies, who can be shown to have E1 character, and a resonance-like structure around 7 MeV, which contains E1 and M1 parts due to the energetic overlap of the PDR and the spinflip-M1 resonance.
The prominents structure peaking at 13 MeV represents the IVGDR.

A separation of E1/M1 cross sections and contributions from other multipoles is possible with a multipole decomposition analysis (MDA) of the angular distributions \cite{pol12,kru15,mar17}.
The measurement of spin transfer observables with a polarized beam provides a separation of spinflip and non-spinflip cross sections \cite{tam11,has15,mar17}, which can be related to E1 and M1 components by the different reaction mechanisms.   
Good agreement is found between these completely independent methods.   
 
The E1 cross sections can be converted to B(E1) strengths, respectively photabsorption cross sections, with the virtual photon method \cite{ber88}.
The bottom part of Fig.~\ref{fig2} shows a comparison of the deduced B(E1) strength distribution in $^{208}$Pb  with data from $(\gamma,\gamma^\prime)$ and (n,$\gamma$) reactions \cite{rye02,sch10,koe87} (l.h.s.) and photoabsorption experiments \cite{vey70,sch88} in the giant resonance region (r.h.s.).
Excellent agreement is obtained \cite{tam11}.
 
The M1 cross sections  can be converted to spin-M1 matrix elements  with the ``unit cross section method'' originally developed to extract the analog GT strength from charge-exchange reactions \cite{ich06}.
Assuming that orbital contributions to the total M1 strength are negligble \cite{hey10} one can extract electromagnetic B(M1) strength distributions from the proton scattering data  \cite{bir16,mat17}.    
Im the case of $^{208}$Pb the M1 contribution to the GSF is small, not exceeding 10\% at the maximum of the resonance.

\begin{figure}[t]
\begin{center}
\includegraphics[width=8cm]{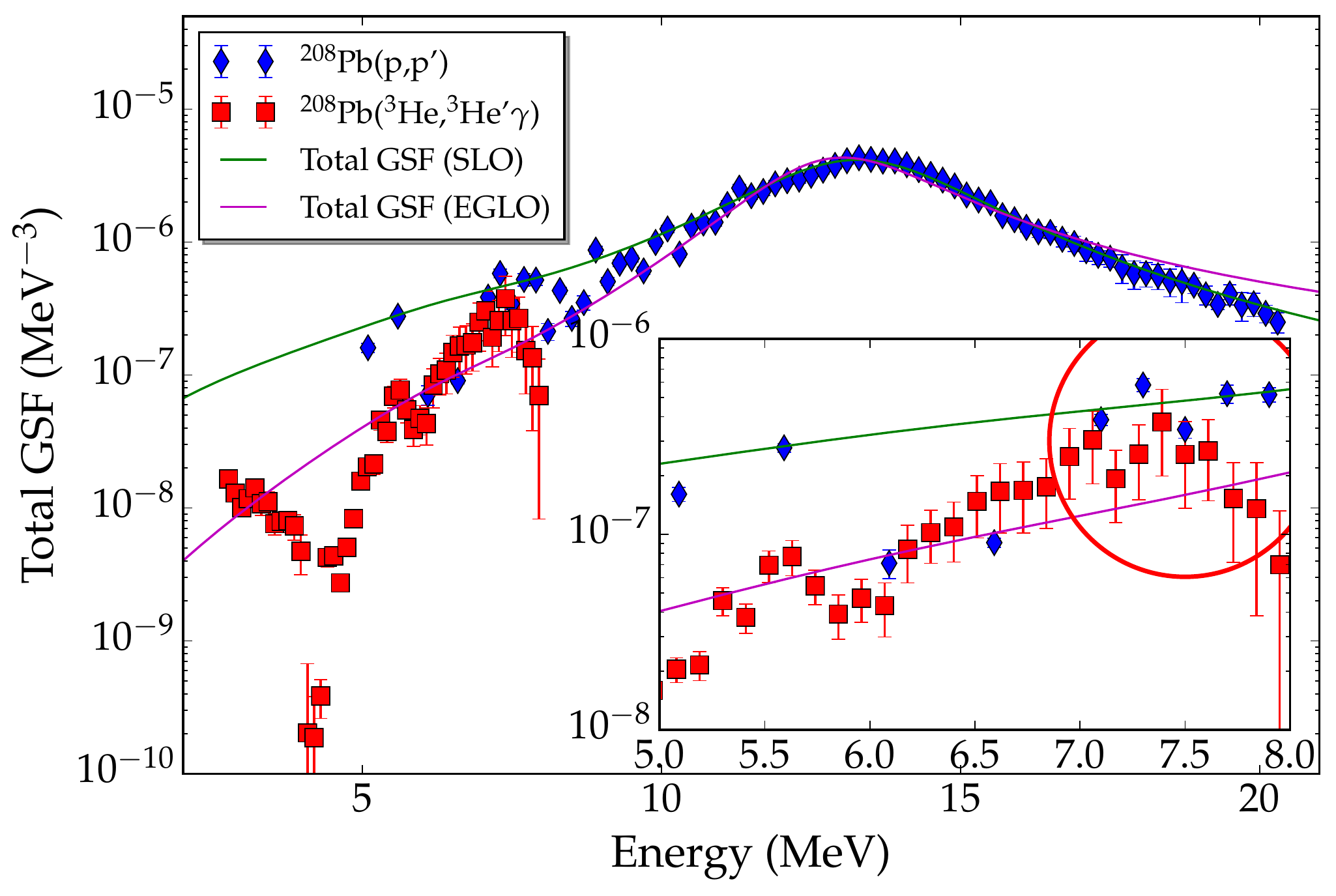}
\caption{\label{fig3}
 GSF deduced from the $^{208}$Pb(p,p$^\prime$) data \cite{tam11,pol12} in comparison with results from an Oslo-type experiment \cite{sye09}.
Reprinted with permission from \url{https://doi.org/10.1103/PhysRevC.94.054313}.
 \copyright 2016 by the American Physical Society.
 }
 \end{center}
\end{figure}

Figure \ref{fig3} presents the GSF deduced from the $^{208}$Pb(p,p$^\prime$) data \cite{bas16} in comparison to results from an Oslo experiment \cite{sye09}.
The inlet shows an extension of the low-energy region, where both experiments overlap.  
The comparison of the present GSF derived from ground-state absorption with the Oslo results shows larger values in the PDR energy region, where both data sets overlap.
However, the fluctuations of the GSF are very strong due to the anomalously small level densities in the closed-shell nucleus $^{208}$Pb, which prevents conclusions on a possible violation of the BA hypothesis in the PDR energy region.

Fluctuations of the cross sections in the energy region of the IVGDR are observed in the high-resolution (p,p$^\prime$) experiments (cf.\ Fig.~\ref{fig2}).
They can be related to the density of $J^\pi = 1^-$ states.
The LD is extracted with a fluctuation analysis decribed e.g.\ in Refs.~\cite{pol14,kal06,kal07,usm11a}.
A prerequiste of the method is a separation of the cross sections populating the IVGDR from other contributions.
In the present case this is achieved by using the MDA results. 

A quantitative description of the fluctuations is given by the autocorrelation function
\begin{equation}
    \label{eq:autocorexp}
        C\left( \epsilon  \right) =
        \frac{{\left\langle {d\left( {E_x } \right) \cdot d\left( {E_x  +
        \epsilon } \right)} \right\rangle }}{{\left\langle {d\left( {E_x }
        \right)} \right\rangle  \cdot \left\langle {d\left( {E_x  +
        \epsilon } \right)} \right\rangle
        }}\;.
\end{equation}
The value $C(\epsilon = 0) - 1$ is nothing but the variance of $d(E_x)$
\begin{equation}
    \label{eq:autocorvar}
        C\left( {\epsilon  = 0} \right) - 1 = \frac{{\left\langle {d^2
        \left( {E_x } \right)} \right\rangle  - \left\langle {d\left( {E_x
        } \right)} \right\rangle ^2 }}{{\left\langle {d\left( {E_x }
        \right)} \right\rangle ^2 }}\;.
\end{equation}
According to Ref.~\cite{jon76}, this experimental autocorrelation function can be approximated by the expression
\begin{equation}
\label{eq:autocorrtheo}
C(\epsilon) - 1 =  \frac{\alpha \cdot \langle \mbox{D} \rangle}{2 \Delta E \sqrt{\pi}} \times f(\sigma,\sigma_>),
\end{equation}
where the function $f$ depends on the chosen parameters (folding widths $\sigma, \sigma_>$) only.
$\alpha$ is the sum of the normalized variances of the assumed spacing and transition width distributions.
If only transitions with the same quantum numbers ($J^\pi =1^-$ in the present case) contribute to the spectrum, it can be directly determined as the sum of the variances of the Wigner and Porter-Thomas distribution, respectively and the mean level spacing $\langle D \rangle$ and LD $\rho(E)=1/\langle D\rangle$ can be extracted from Eq.~(\ref{eq:autocorvar}).

\begin{figure}[t]
\begin{center}
\includegraphics[width=8cm]{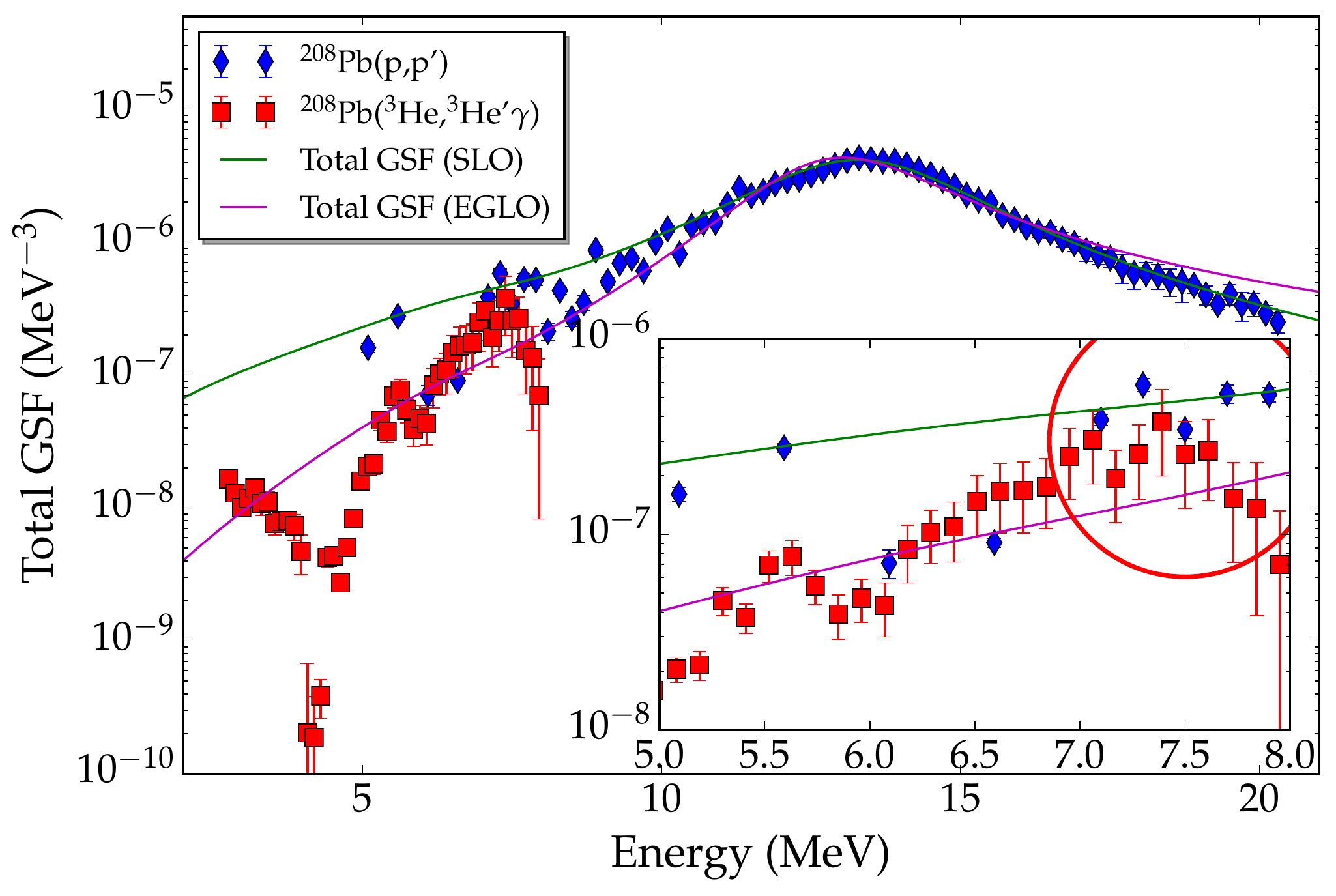}
\caption{\label{fig4}
Total LD in $^{208}$Pb from the ($p,p'$) data \cite{tam11,pol12} in comparison with results from an Oslo-type experiment \cite{sye09}.
Reprinted with permission from \url{https://doi.org/10.1103/PhysRevC.94.054313}. 
\copyright 2016 by the American Physical Society.}
\end{center}
\end{figure}

In order to compare with the results from the Oslo experiment, the $1^-$ LD needs to be converted to a total LD.
The spin distribution is calculated with the aid of systematic backshifted Fermi-gas model (BSFGM) parameterizations and their variation is taken as a measure of the systematic uncertainty of the procedure (for details see Ref.~\cite{bas16}).
Figure \ref{fig4} displays the resulting LD in the region 9.5-12.5 MeV (blue diamonds) together with resuts fo the Oslo experiment at lower energies (red squares) \cite{sye09} and the data point at neutron threshold from neutron capture \cite{cap09}.  
Several BSFGM results are shown as solid, dashed and dotted lines, respectively. 
The RIPL-3 parameterization \cite{cap09} provides a very satisfactory description of all experimental data indicating that the decomposition into GSF and LD in the Oslo method is essentially correct. 

\begin{figure}
\begin{center}
\includegraphics[width=12cm]{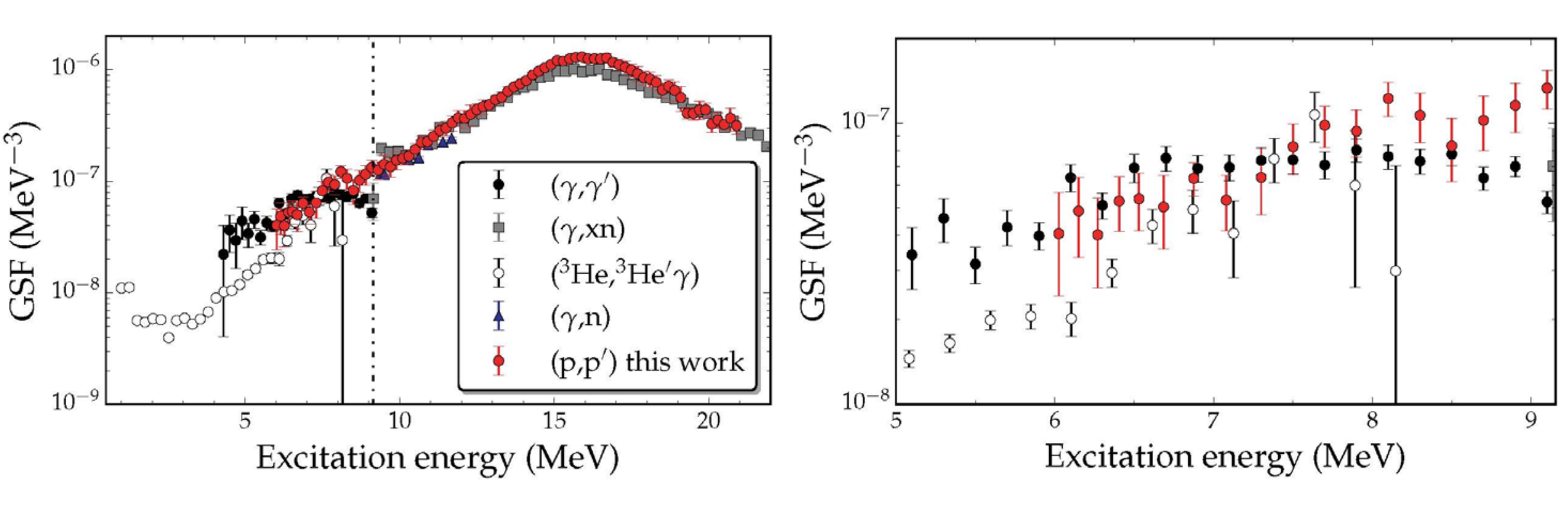}
\caption{\label{fig5}
GSF of $^{96}$Mo from the the ($p,p'$) data (red circles) compared with ($^3{\rm He},^3{\rm He}^\prime \gamma$) \cite{gut05,lar10} (open circles) and ($\gamma,\gamma^\prime$) data including a statistical model correction for unobserved branching ratios \cite{rus09} (black circles).
See Ref.~\cite{mar17}.}
\end{center}
\end{figure}

Another study of this type was performed for $^{96}$Mo \cite{mar17}, a considerably deformed nucleus with LDs high enough to permit a comparison with the GSF from a decay experiment averaging over appropriate energy intervals. 
The choice of  $^{96}$Mo was motivated by the large discrepancies of GSFs derived from Oslo \cite{gut05,lar10} and NRF \cite{rus09} experiments.
The l.h.s.\ of Fig.~\ref{fig5} summarizes the available GSF data.
The energy region below neutron threshold is expanded on the r.h.s.\ showing the results from
the Oslo (open circles), the NRF (black circles), and the (p,p$^\prime$) experiment (red circles).
For $\gamma$ energies between 6 an 8 MeV covered by all experments, the GSF deduced from Coulomb excitation lies between the the two other results but overall agrees better with the Oslo result (for details see Ref.~\cite{mar17}). 

\begin{figure}
\begin{center}
\includegraphics[width=8cm]{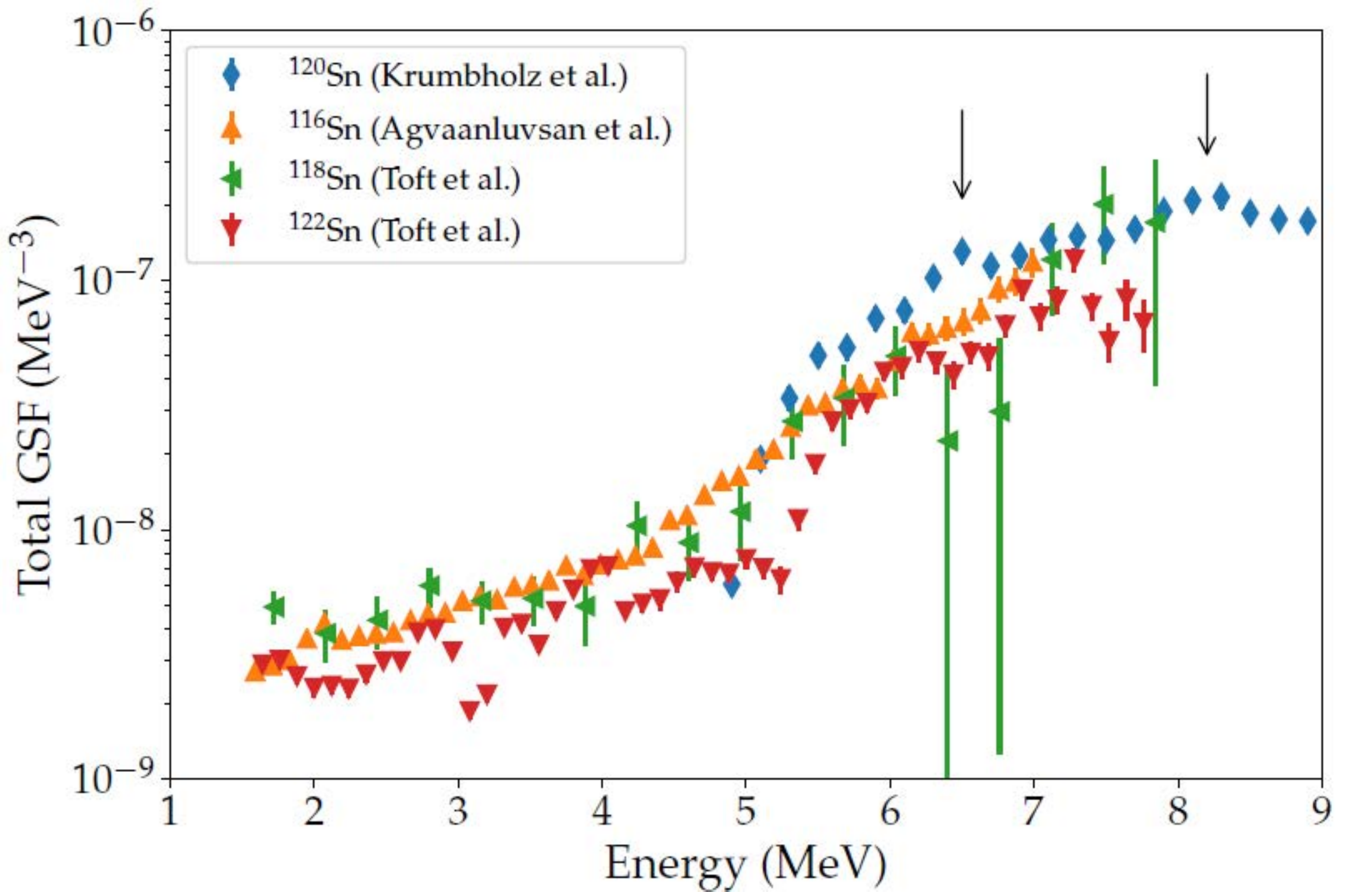}
\caption{\label{fig6}
GSF of $^{120}$Sn in the energy region from 5 to 9 MeV from the the ($p,p'$) data \cite{kru15,has15} in comparison with Oslo-type results for $^{116}$Sn \cite{agv09} and $^{118,122}$Sn \cite{tof11}.}
\end{center}
\end{figure}

Finally, we have extracted the GSF of $^{120}$Sn from the data described in Refs.~\cite{kru15,has15}, again including the M1 part due to the spinflip resonance. 
In the GDR region fair agreement with previous experiments is obtained \cite{has15}.
The energy region below neutron threshold is displayed in Fig.~\ref{fig6} and exhibits two pronounced resonance-like structures around 6.5 and 8 MeV indicated by arrows.
Data from an Oslo-type experiment are not available for $^{120}$Sn, however, the neighboring even-even Sn isotopes 116 \cite{agv09} and 118,122 \cite{tof11} have been studied.
Since the low-energy structure is known to change little across the stable even-even Sn istopes one can also expect that changes of the GSF are limited (although the PDR is expected to have some dependence on neutron excess \cite{sav13}).

For $\gamma$ energies from 5 to about 7.5 MeV covered by both types of experiments one finds reasonable agreement at the lower and upper end of the interval.
In contrast, the Oslo data show a smooth energy dependence and no resonance-like structure around 6.5 MeV pointing to a violation of the BA hypothesis.
It should be noted that this bump is systematically seen in $0^\circ$ (p,p$^\prime$) cross sections for all stable even-even Sn isotopes \cite{bas19} and has also been observed in $^{124}$Sn with isoscalar probes \cite{end10,pel14}.

\section{Concluding remarks}

The generalized BA hypothesis is a crucial assumption for the application of statistical nuclear reaction theory with photons in the entrance or exit channel.
Of particular importance is the question whether data from g.s.\ absorption experiments represent the GSF in the (quasi)continuum region.
While its validity is fairly well established above neutron threshold in medium-mass and heavy nuclei, the situation is less clear at lower $\gamma$ energies when comparing decay and absoprtion experments.
There are clear violations like the LEE and the larger scissors mode strength in the decay.
For the (PDR + spinflip M1) energy region there are conflicting results.

The present contribution discusses a new approach to extract the GSF (including the spin-M1 part) from (p,p$^\prime$) scattering at energies of a few hundred MeV and at very forward angles.   
This method directly measures the g.s.\ decay width and avoids the problems of NRF data, where one needs to correct for unknown branching ratios to excited states. 
When performed with high energy resolution, such data not only provide the GSF but also the LD, thus permitting an important test of the assumptions made in Oslo-type experiments for their decomposition.
So far, three cases have been analyzed.
The study of $^{208}$Pb remains inconclusive because the anomalously low LD leads to large intensity fluctuations \cite{bas16}.
For $^{96}$Mo consistency within the experimental uncertainties is found \cite{mar17}.
The results in $^{120}$Sn point to a violation of the BA hypothesis \cite{bas19}.
Clearly, a more systematic study is needed -- e.g.\ on the role of deformation --  and emphasis should be put to establish more cases, where GSF and LD from Oslo-type and (p,p$^\prime$) experiments (as well as the LD from neutron capture) can be compared.       

I thank S.~Bassauer for his contribution to the analysis of the present results and A.~Tamii and the collaborators at RCNP for the excellent experiments.
This work was funded by the Deutsche Forschungsgemeinschaft (DFG, German Research Foundation) -- Projektnummer 279384907 -- SFB 1245.  

%
%


\begin{thebibliography}{abc99x}
%

\bibitem{arn07}
M. Arnould, S. Goriely, and K. Takahashi, Phys. Rep. {\bf 450}, 97 (2007).

\bibitem{cha11}
M.B. Chadwick {\it et al.}, Nucl. Data Sheets {\bf 112}, 2887 (2011).

\bibitem{sal11}
M. Salvatore and G. Palmiotti, Prog. Part. Nucl. Phys. {\bf 66}, 144 (2011).

\bibitem{wie12}
M. Wiescher, F. K\" {a}ppeler, and K. Langanke, Annu. Rev. Astron. Astrophys. {\bf 50}, 165 (2012).

\bibitem{bri55}
D.M. Brink, Ph.D. thesis, Oxford University (1955).

\bibitem{axe62}
P. Axel, Phys. Rev. {\bf 126}, 671 (1962).

\bibitem{bbb98}
P.F. Bortignon, A.Bracco, and R.A. Broglia, {\it Giant Resonances: Nuclear Structure at Finite Temperature} (Harwood Academic, Amsterdam, 1998).

\bibitem{joh15}
C. W. Johnson, Phys. Lett. B {\bf 750}, 72 (2015).

\bibitem{hun17}
N. Quang Hung, N. Dinh Dang, and L.T. Quynh Huong, Phys. Rev. Lett. {\bf 118}, 022502 (2017).

\bibitem{sch00}
A. Schiller {\it et al.},
Nucl. Instrum. Methods  A {\bf 447}, 498 (2000).

\bibitem{gut16}
M. Guttormsen {\em et al.},
Phys. Rev. Lett.  {\bf 116}, 012502 (2016).

\bibitem{lar17}
A.C. Larsen {\em et al.},
J. Phys.  G {\bf 44}, 064005 (2017).

\bibitem{boh84}
D. Bohle {\em et al.},
Phys. Lett. B {\bf 137}, 27 (1984).

\bibitem{hey10}
K. Heyde, P. von Neumann-Cosel, and A. Richter, 
%
Rev. Mod. Phys. {\bf 82}, 2365 (2010).

\bibitem{gut12}
M. Guttormsen {\it et al.},  Phys. Rev. Lett. {\bf 109}, 162503 (2012).

\bibitem{ang16}
C.T. Angell {\em et al.},
%
Phys. Rev. Lett. {\bf 117}, 142501 (2016).

\bibitem{voi04}
A. Voinov {\em et al.},
Phys. Rev. Lett. {\bf 93}, 142504 (2004).

\bibitem{rus09}
G. Rusev {\it et al.}, Phys. Rev. C {\bf 79}, 061302 (2009).

\bibitem{rom15}
C. Romig {\it et al.}, Phys. Lett. B {\bf 744}, 369 (2015).

\bibitem{loe16}
B. L\"oher {\it et al.}, Phys. Lett. B {\bf 756}, 72 (2016).

\bibitem{isa19}
J. Isaak {\it et al.}, Phys. Lett. B {\bf 788}, 225 (2019).

\bibitem{ang12}
C.T. Angell {\em et al.},
Phys. Rev. C {\bf 86}, 051302(R) (2012).

\bibitem{ang15}
Erratum to Ref.~\cite{ang12}, Phys. Rev. C {\bf 91}, 039901(E) (2015).

\bibitem{tam11}
A. Tamii {\it et al.}, 
%
Phys. Rev. Lett. {\bf 107}, 062502 (2011).

\bibitem{pol12}
I. Poltoratska {\it et al.},
%
Phys. Rev. C {\bf 85}, 041304(R) (2012).

\bibitem{kru15}
A.M. Krumbholz {\it et al.},
%
Phys. Lett. B {\bf 744}, 7 (2015).

\bibitem{has15}
T. Hashimoto {\it et al.},
%
Phys. Rev. C {\bf 92}, 031305(R) (2015).

\bibitem{bir17}
J. Birkhan {\it et al.},
%
Phys. Rev. Lett. {\bf 118}, 252501 (2017).

\bibitem{mar17}
D. Martin {\em et al.},
%
Phys. Rev. Lett. {\bf 119}, 182503 (2017).

%

\bibitem{bir16}
J. Birkhan {\em et al.},
Phys. Rev. C {\bf 93}, 041302(R) (2016).

\bibitem{pol14}
I. Poltoratska {\it et al.},
%
Phys. Rev. C {\bf 89}, 054322 (2014).

\bibitem{bas16}
S. Bassauer, P. von Neumann-Cosel, and A. Tamii,
%
Phys. Rev. C {\bf 94}, 054313 (2016).

\bibitem{tam09}
A. Tamii {\em et al.},
%
Nucl. Instrum. Methods A {\bf 605}, 3 (2009).

\bibitem{ber88}
%
C.A. Bertulani and G. Baur,
%
Phys. Rep. {\bf 163}, 299 (1988).

\bibitem{rye02}
N. Ryezayeva {\em et al.},
%
Phys. Rev. Lett. {\bf 89}, 272502 (2002).

\bibitem{sch10}
R. Schwengner {\em et al.},
Phys. Rev. C {\bf 81}, 054315 (2010).

\bibitem{koe87}
R. K\"ohler {\em et al.},
%
Phys. Rev. C {\bf 35}, 1646 (1987).

\bibitem{vey70}
A. Veyssiere {\em et al.},
%
Nucl. Phys. {\bf A159}, 561 (1970).

\bibitem{sch88}
K.P. Schelhaas {\em et al.},
%
Nucl. Phys. {\bf A489}, 189 (1988).

\bibitem{ich06}
M. Ichimura, H. Sakai, and T. Wakasa,
%
Prog. Part. Nucl. Phys. {\bf 56}, 446 (2006).

\bibitem{mat17}
M. Mathy {\it et al.},
%
Phys. Rev. C {\bf 95}, 054316 (2017)

\bibitem{sye09}
N.U.H. Syed {\em et al.},
%
Phys.~Rev.~C {\bf 79}, 024316 (2009).

\bibitem{kal06}
Y. Kalmykov {\it et al.},
%
Phys. Rev. Lett. \textbf{96}, 012502 (2006).

\bibitem{kal07}
Y. Kalmykov {\em et al.},
%
Phys. Rev. Lett. \textbf{99}, 202502 (2007).

\bibitem{usm11a}
I. Usman {\it et al.},
%
Phys. Rev. C {\bf 84}, 054322 (2011).

\bibitem{jon76}
B. Jonson {\it et al.},
%
CERN report {\bf 76--13}, 277 (1976).

\bibitem{cap09}
R. Capote {\it et al.}, 
%
Nucl. Data Sheets {\bf 110}, 3107 (2009).

\bibitem{gut05}
M. Guttormsen {\it et al.}, Phys. Rev. C {\bf 71}, 044307 (2005).

\bibitem{lar10}
A.C. Larsen and S. Goriely, Phys. Rev. C {\bf 82}, 014318 (2010).


\bibitem{agv09}
U. Agvaanluvsan {\it et al.},
%
Phys. Rev. C {\bf 79}, 014320 (2009).

\bibitem{tof11}
H.K. Toft {\it et al.},
%
Phys. Rev. C {\bf 83}, 044320 (2011).

\bibitem{sav13}
D. Savran, T. Aumann, and A. Zilges,
%
Prog. Part. Nucl. Phys. {\bf 70}, 210 (2013).

\bibitem{bas19}
S. Bassauer {\it et al.}, to be published.

\bibitem{end10}
J. Endres {\it et al.},
%
Phys. Rev. Lett. {\bf 105}, 212503 (2010).
 
\bibitem{pel14}
L. Pellegri {\it et al.},
%
Phys. Lett. B {\bf 738}, 519 (2014).

\end{thebibliography}
\end{document}